\documentclass{iopart}
\usepackage{graphicx}

\usepackage{bm}
\usepackage[german,english]{babel}

\begin{document}

\title[Ultracold bosons in 1D harmonic and multi-well traps: QMC vs CPWF approach]{Ultracold bosons in one-dimensional harmonic and multi-well traps: a Quantum Monte Carlo vs a correlated pair approach}

\author{I. Brouzos$^1$, F. K. Diakonos$^2$, P. Schmelcher$^1$}

\address{$^1$ Universit\"at Hamburg, Zentrum f\"ur Optische Quantentechnologien, Luruper Chaussee 149, 22761 Hamburg,  Germany}
\address{$^2$Department of Physics, University of Athens, GR-15771 Athens, Greece }

\ead{\mailto{ibrouzos@physnet.uni-hamburg.de}, \mailto{fdiakono@phys.uoa.gr}, \mailto{pschmelc@physnet.uni-hamburg.de}}

\begin{abstract}
We study the crossover of a finite one-dimensional (1D) bosonic ensemble from weak to strong interactions in harmonic traps and multi-well potentials. Although these systems are very common in experimental setups and have been studied theoretically, an analytical description is lacking. We perform Diffusion Quantum Monte Carlo calculations which we show to be in good agreement with results from analytical functions that we construct to describe these systems. For the harmonic trap we use the correlated-pair wave function which we introduced in \cite{brouzos} considering here much larger atom numbers, going beyond the few-body ensembles studied  in \cite{brouzos}. We also investigate double and triple wells, changing correspondingly the uncorrelated part of the Ansatz to describe efficiently the single-particle behaviour. On-site effects  beyond mean-field and standard Bose-Hubbard calculations that appear in densities being captured by our analytical functions are explored.  
\end{abstract}

\submitto{JPB}
\maketitle

\section{Introduction}

One-dimensional many-body problems have a long tradition in theoretical physics since a substantial number of them allows for an analytical description mainly by employing the Bethe-Ansatz. A seminal example is the Lieb-Liniger gas \cite{lieb} for bosons interacting via a contact potential of arbitrary strength in the absence of an external potential with periodic boundary conditions  (and extendable also to hard-wall boundaries \cite{hao}). The existence of an exact solution for such a general many-body problem has lead to further studies, also trying to extend this idea to the presence of  external traps. One other important peculiarity of 1D gases is the existence of the so-called Tonks-Girardeau gas \cite{girardeau}, where hard-core bosons interacting with infinite strength can be mapped to identical non-interacting fermions, since the infinite repulsion is mimicking the Pauli-exclusion principle. 

In the last decade, the excitement about 1D physics has received significant attention by corresponding experimental realizations, since the preparation of cold gases in quasi-1D traps with strong transversal confinement has become possible. The Tonks-Girardeau gas \cite{kinoshita} and also its counterpart for the strongly attractive excited state, the super Tonks-Girardeau gas \cite{haller} have been observed, and numerous other 1D phenomena like the pinning transition \cite{pinning} with a weak optical lattice have been explored, extending also to nonequilibrium quantum dynamics and thermalization processes \cite{trotzky}.  All these would not be possible without the high degree of experimental control of the trapping geometry via external fields \cite{bloch} as well as the interaction of ultracold atomic ensembles via Feshbach \cite{chin} or confinement induced resonances \cite{olshanii}, a tool specific to quasi-1D waveguide-like systems. The precision of measurements has been recently extended to single-cite addressing in optical lattices \cite{greiner}, while the deterministic preparation of the system has reached the possibility of loading an exact number of atoms into the trap \cite{selim1, selim2}.

Theoretical studies of 1D systems have been mainly focusing on the uniform and the harmonic trap case. Especially Monte Carlo simulations in configuration space have been performed using trial functions of Jastrow type as a guiding for the diffusion  \cite{qmc}. The common Bijl-Jastrow trial function involves two variational parameters: one for the  two-body interaction part and one for the single-particle part. 
%Our proposal here is to use for the two-body part  a function introduced in \cite{brouzos} which is inspired by the exact solution of the problem of two atoms in the harmonic trap \cite{busch}. 
On the other hand, one-dimensional lattices have been studied typically within the Bose-Hubbard model \cite{bhm}, which is a tight-binding approach and in the usual form cannot capture on-site effects in the wave-function as long as unperturbed Wannier on-site orbitals are used. Numerical approaches which demonstrate localization and delocalization mechanisms, as well as on-site features like fragmentation and multi-particle effective  interactions have been developed \cite{lattice}.

In this work we perform a comparative study between quantum Monte-Carlo calculations and results from analytical wave-functions for the description of ground state energies and correlation properties of bosonic systems in a harmonic and multi-well traps. Our approach initiates from the exact solution of the two-body problem in the trap \cite{busch} and the generalization in terms of a correlated-pair wave-function (CPWF) which we have proposed in \cite{brouzos} in order to cover the few-body (up to 6 atoms) case in the harmonic potential. We extend the calculations we have performed in \cite{brouzos} the many-body regime (up to 50 atoms) in the trap performing a diffusion Monte Carlo (DMC) simulation which provides us with the numerically-exact energy of the system which we compare with the corresponding numerical integration of the CPWF. We also extend the CPWF Ansatz to cover multi-well potentials (double and triple well). For the construction of a suitable parameter-free Ansatz here we use localized Wannier functions of Gaussian profile for the single particle part of the wave function to describe the arrangement of the particles in the multi-well trap, and hypergeometric functions inspired from the two-body problem in the trap to include pair-correlation and on-site effects. A Variational Monte Carlo (VMC) approach allowing for one or two variational parameters has been also employed without substantial differences with respect to the obtained energy values compared to our parameter-free Ansatz. We demonstrate on-site features appearing in multi-well potentials for several observables (one-body density and pair- correlation functions) captured by our analytical functions.

This article is organized as follows: in Section II we introduce our setup, its Hamiltonian, as well as our Ansatz. In Section III, we compare results for the energy from DMC simulations with these from the trial function CPWF. We also show one- and two-body   density functions, focusing on on-site effects. Finally we summarize our results and give an outlook in Section IV.

\section{Setup and Ansatz}

\subsection{Hamiltonian for harmonic and multi-well traps}

For the investigation in particular of 1D systems one should take into account the experimental conditions and their impact on the collisional properties of the atoms. Experimentally, the standard method to create quasi-1D tubes is using a very strong laser field for the transversal directions compared to the lateral one \cite{kinoshita,haller}. This way the trap becomes highly anisotropic with the characteristic transversal length scale  $a_{\perp} \equiv \sqrt {\frac{\hbar}{M \omega_{\perp}}}$ being much smaller than the longitudinal one $a_{\parallel} \equiv \sqrt {\frac{\hbar}{M \omega_{\parallel}}}$  where $\omega_{\perp}$ ($\omega_{\parallel}$) is the transversal (longitudinal) harmonic confinement frequency. Then the transverse degrees of freedom are energetically frozen to their ground states and the effective 1D interaction strength reads \cite{olshanii}:
\begin{equation*}
g_{1D}= \frac{2\hbar^2 a_0}{M a^2_{\perp}} \left(1-\frac{|\zeta(1/2)| a_0}{\sqrt{2} a_{\perp}}\right)^{-1},
\end{equation*} 
where $a_0$ is the 3D s-wave scattering length. 

We will study here two systems. First as a standard benchmark example a 1D harmonic trap where the rescaled 1D N-body Hamiltonian reads:
\begin{equation*}
H_{\mathrm{ho}} = - \frac{1}{2} \sum_{i=1}^N \frac{\partial^2}{\partial x_i^2} + \frac{1}{2} \sum_{i=1}^N x_i^2 + g \sum_{i<j}\delta \left( x_{ij} \right).
\end{equation*}
In $H_{\mathrm{ho}}$ the atoms interact with a contact potential modeled by the Dirac $\delta$-function with $x_{ij}=x_i-x_j$ denoting the relative coordinate of the i-th and j-th atom. The lengths and the energies are scaled by $a_{\parallel}$ and $\hbar\omega_{\parallel}$, respectively. Thus, the single remaining parameter is the rescaled interaction strength  $g=\frac{g_{1D}} {\hbar \omega_{\parallel} a_{\parallel}}$ which is controlled either by tuning $a_0$ via magnetic Feshbach resonances or $a_{\perp}$ by modifying the transversal confinement. This system has been studied for few atoms also in \cite{brouzos,deuretzbacher,sascha}. Here we extend the analytical approach we have introduced in \cite{brouzos} to a setup involving a larger number of atoms. 

Apart from this single-site prototype system, we are also interested here in finite multi-well systems. These can be prepared experimentally e.g. by using more than one counter-propagating laser beams (standard optical lattice technique) and forming a superlattice of copies of small finite lattices \cite{trotzky}, or from the beam waist which always produces an approximate harmonic confinement resulting in a concentration of the cloud density in the few central wells of the potential. The standard optical lattice shows a potential profile of the form $V_0 \sin^2 \left(\frac{\pi x_i}{2 \alpha} +\phi \right)$ with a superimposed harmonic confinement. The rescaled 1D Hamiltonian reads: 
\begin{equation*}
H_{\mathrm{lat}} = - \frac{1}{2} \sum_{i=1}^N \frac{\partial^2}{\partial  x_i^2} + \sum_{i=1}^N V_0 \sin^2 \left(\frac{\pi x_i}{2}+\phi \right) + g \sum_{i<j}\delta \left( x_{ij} \right)
\end{equation*}
where the lengths are scaled by the lattice constant $\alpha$ and the energies by the recoil energy $E_R=\frac{\hbar^2}{m \alpha^2}$. We further confine this infinite lattice system to a finite region $L~\in~[-5/2,5/2]$ (with $\phi=\pi/2$) and $L~\in~[-7/2,7/2]$  (with $\phi=0$) imposing hard-wall boundary conditions at the edges, such that two and three wells are isolated, respectively. This way we encounter simple models of finite multi-well systems, a double and a triple well potential. The Hamiltonian is characterized by two parameters: the rescaled interaction strength $g=\frac{g_{1D}}{E_R \alpha}$ and the  potential depth $V_0$ (in units of $E_R$). We will consider here rather deep lattices ($V_0=40$) such that the on-site effects can be demonstrated and also a tight-binding approximation with well-localized Gaussian funtions at each well, which we will introduce next, can be considered as a good approach.

\subsection{Ansatz} 

In \cite{brouzos} we have proposed a correlated pair wave function (CPWF) inspired from the idea that the analytical solution for a single pair in the harmonic trap can be extended to the many-body system by taking products of exact two-body functions. The contact interaction may be then adequately addressed, if the discontinuity that it causes is imposed on each pair of atoms in the ensemble, in a similar way as for a single pair. The CPWF reproduces the two exactly solvable limits of zero and infinite interaction strength (Tonks- Girardeau gas) for an arbitrary number of atoms in the harmonic trap. We have already indicated  in \cite{brouzos} that the CPWF can be used for Monte Carlo simulations since it has the form of a Bijl-Jastrow function, but with a different approach for the two-body part than the typically used trigonometric functions \cite{qmc}. In this paper we further develop this idea applying it also to  higher number of particles, and we extend it to the case of a finite-size multi-well potential. 

For a VMC calculation it is crucial to have a good functional form of the trial function  which can then be used also as a guiding function for DMC simulations driving the walkers to positions where, for physical reasons, it is more possible for the particles to be placed. For the implementation of DMC we follow similar prescriptions like those explained in \cite{qmc}. The typical trial-guiding function used in these approaches use are of the form of Bijl-Jastrow functions, with a single particle part (SPP) which accounts for the form of the external potential and an interaction part (IP) which accounts for the collision. For a 1D harmonic trap the standard form reads: 

\begin{equation}
\label{stand_trial}
 \Psi_{T} = \underbrace{\prod_i e^{-\beta x_i^2/2}}_{\psi_{SPP}} \underbrace{\prod_{i<j} \cos \left[ k \left(|x_{ij}|- \frac{L}{2} \right) \right]}_{\psi_{IP}}
\end{equation}

The form of the SPP corresponds obviously to the ground state of a single particle in a harmonic trap (setting also the presumably variational parameter $\beta=1$). For the IP the standard Bijl-Jastrow type of functions are used, and an interesting remark is that the form  $ \cos \left[ k \left(|x_{ij}|- \frac{L}{2} \right) \right]$ is actually the solution of the two-particle Lieb-Liniger gas \cite{lieb} (in homogeneous space) of length $L$ with $k$ determined from the boundary condition (or Bethe condition as it is often called) at the contact point $x_{ij}=0$ [discontinuity of the derivative:  $2 \psi_{ij}'(0)= g \psi_{ij}(0) \Rightarrow g=k \tan \left( \frac{kL}{2} \right)$]. For long range interparticle distances $|x_{ij}|>L$ one takes a constant value for the IP. Notice that this approach contains one variational parameter $L$ (if we consider $\beta$ fixed) which should be optimized.

Seen from this perspective, our approach in \cite{brouzos}  implies the replacement of the IP with the hypergeometric functions $U \left( -\frac{\nu}{2}, \frac{1}{2}, \frac{x_{ij}^2}{2} \right)$ \cite{abramowitz}, which apart from an exponential term (which combines with the center of mass motion), are the analytical solution of the relative motion of two particles in a trap $\left( -\frac{d^2}{ d x_{ij}^2}+ \frac{x_{ij}^2}{4} -e_{ij}  \right) \psi \left( x_{ij} \right)=0$, where $e_{ij}=\nu+\frac{1}{2}$. The parameter $\nu$ is a generalized quantum number which can take real values (and not only integer as in the case of a harmonic trap without interaction) and is determined according to the boundary condition at the collision point  $x_{ij}=0$: 
\begin{equation*}
\label{bc}
g=-\frac{2^\frac{3}{2}\Gamma\left(\frac{1-\nu}{2}\right)}{\Gamma\left(\frac{-\nu}{2}\right)}
\end{equation*}  

Our Ansatz  for the wave-function in the harmonic trap then reads: 
\begin{equation}
\label{cpwf}
 \Psi_{T} = \prod_i e^{-\beta x_i^2/2} \prod_{i<j} U \left( -\frac{\nu}{2},\frac{1}{2}, \frac{x_{ij}^2}{2} \right)
\end{equation}
and has no variational parameter in the IP, while the one in the SPP can be fixed to $\beta=1$ according to the ground state of the Hamiltonian $H_{ho}$ with $g=0$ (being simultaneously correct for the Tonks-Girardeau gas limit $g \to \infty$, see \cite{brouzos,girardeau1}).  This form Eq.~\ref{cpwf} is equivalent to the one appearing in \cite{brouzos} since the relation $ D_{\nu}(x)= 2^{\frac{\nu}{2}}e^{-\frac{x^2}{4}}U(-\frac{\nu}{2},\frac{1}{2},\frac{x^2}{2})$ holds for $x>0$ \cite{abramowitz}. 

A remark on the two approaches is in order here: the functional form of the standard approach [Eq. (\ref{stand_trial})] can come very close to the one proposed here [Eq. (\ref{cpwf})], since it contains a variational parameter $L$ which when optimized leads to a functional form for the IP close to the hypergeometric functions. The advantage of the proposed approach is that it can be used without any free parameter. Still one can insert one variational parameter in the last argument of the hypergeometric function and use it variationally too. Nevertheless, our aim in this work is rather to propose a function that can describe correctly the behaviour without free parameters. Since both approaches are employing in some sense the two-particle solution in the continuum [Eq. (\ref{stand_trial})] and in the trap [Eq. (\ref{cpwf})], they offer a physical picture which generalizes the two-body problem to the many-body one. 

A different SPP part is needed for the case of a deep finite multi-well potential. In the case of a perturbative lattice one could take the same approach or probably slightly [Eq. (\ref{cpwf})] modified by an inverse lattice term for the SPP like $1-\gamma \sin^2(\pi x_i/2+\phi)$. But for a deep lattice $V_0>4 (E_R)$ a radical change of the SPP should be employed in terms of the tight-binding model which is valid for this case, since next to next-neighbor tunneling is at least one order of magnitude smaller than nearest-neighbor tunneling. In a general sense one can write an expansion of Gaussian functions (as an approximation to the localized Wannier functions) localized at the position $x_j$ of each well $j$:
\begin{equation}
 \label{spp_lat}
\psi_{SPP} \left( x_i \right)= \sum_j f_j e^{-\beta (x-x_j)^2}
\end{equation}
The parameter $\beta$ is related to the effective confinement frequency of each well (considered as a harmonic oscillator), and in our case can be set to $\beta=\sqrt{V_0}$. The coefficients $f_j$ are actually dependent on the interaction between the atoms, and  in the general case can be derived from a Bose-Hubbard model calculation. For weak interaction strengths the atoms tend to be more in the center of the potential (for harmonic or hard-wall confinement) while increasing the interaction strength leads us to the Mott-insulator phase where the particles are essentially equally populating all wells (or forming wedding-cake structures of several Mott shells) \cite{bhm}.  In the case of an optical lattice plus a parabolic confinement the tight-binding model for the single-particle problem has been solved in \cite{rey} by means of Mathieu functions which offer the coefficients for an expansion in terms of  localized functions. We note here that this can be also considered as a parameter free Ansatz for the SPP provided that we use the form in Eq. (\ref{spp_lat}) with a fixed $\beta=\sqrt{V_0}$ \cite{note}.  Yet this will not cover the full range from weak to strong interactions since the distribution of weights will change with increasing interaction strength. We take a detour from this by focusing on interaction effects that appear on-site like those shown in \cite{lattice} and not on the redistribution of the particles from the superfluid to the Mott insulator state.  We will examine here cases with equal population in each lattice site or obtain this equal distribution for a perturbatively weak interaction strength. We will therefore use equal coefficients ($f_j=1$ for all $j$) in the expansion Eq. (\ref{spp_lat}) constructing thus a direct generalization of our Ansatz for lattices (for periodic lattices this choice is actually the only reasonable). In the next section we will examine the adequacy of such an Ansatz to capture the properties of the energy and other observables.

\begin{figure*}
\includegraphics[width=12.0 cm,height=8.0cm]{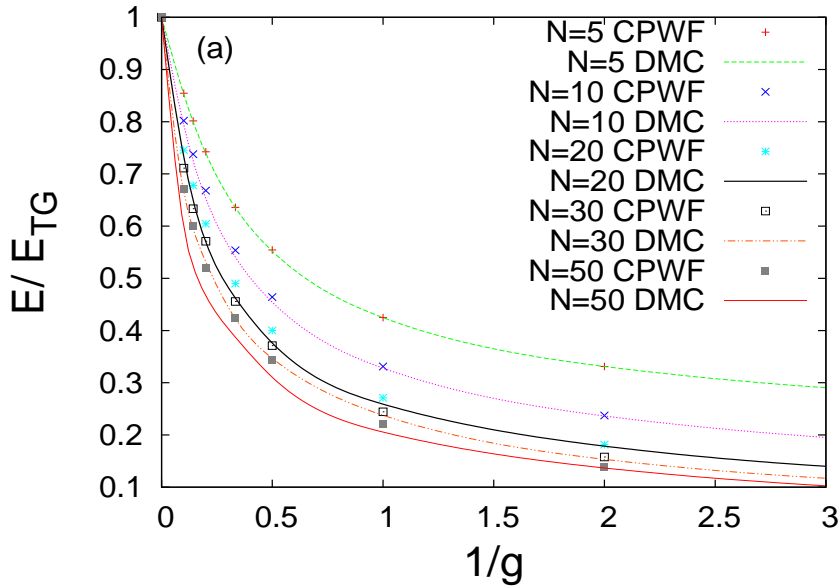}
\caption{ Ground state energies ($E$) divided by the corresponding value of the Tonks-Girardeau limit ($E_{TG}$)  in the harmonic trap as a function of the inverse interaction strength ($g=g_{1D}/\hbar \omega_{\parallel} \alpha_{\parallel}$). Several cases with respect to the number of atoms are shown, comparing results obtained by DMC and numerical integration of the analytical formula CPWF.}
\label{ht}
\end{figure*}

\section{Results} 

\subsection{Harmonic Trap}

In \cite{brouzos} we have shown that for the harmonic trap the energy calculated from the CPWF (Eq. \ref{cpwf}) is in good agreement with the one obtained from numerical calculations for a few atoms. Exact diagonalisation and Multi-Configurational Time-Dependent Hartree, both reliable for a few degrees of freedom, have been used there for the comparisons. In this work we employ VMC and DMC methods to calculate the energy for more atoms up to the standard occupation of each 1D tube in experimental setups varying from 10-50 atoms.  The DMC method \cite{qmc}, is reliable for extracting the energy of a many-body system. 

In Fig. \ref{ht} we show the results for the energy of the ground state as a function of the interaction strength $g$ for several numbers of particles in the trap. Shown is actually the ground state energy of the  interacting many-body system divided by the energy corresponding to the Tonks-Girardeau limit (which is equal to that of the corresponding system of identical fermions $E_{TG}=N^2/2$ \cite{girardeau,girardeau1,brouzos}) as a function  of the inverse of the interaction strength $1/g$ (which in fact is proportional to the effective 1D scattering length $a_{1D} \propto -1/g$ \cite{olshanii}). The uppermost curve and points for $N=5$ atoms (as well as for less atoms-not shown) obtained here from the DMC calculation  and numerical integrations using the functional form of the CPWF Ansatz respectively,  confirm the statements in \cite{brouzos} that the CPWF describes accurately not only qualitatively but also quantitatively the crossover from weak to strong interactions. The agreement is very good allover this crossover, and very accurate especially for very weak $g\to 0$ and very strong interactions close to the resonance (TG limit) $g \to \infty$. As we have already shown in \cite{brouzos} the correlated-pair wave function represents the exact ground state of the system for the non-interacting and TG limit for arbitrary number of atoms. 

For $N=10$ or less (see Fig. \ref{ht}) our Ansatz is still accurate up to $2\%$ error, and can be even improved (even below $1\%$) by VMC letting the parameter $\beta$  of Eq.~\ref{cpwf} being optimized  (this usually leads to lower values of $\beta$). For an increasing number of atoms (compare curves and points in Fig. \ref{ht} for $N=20,30,50$) the intermediate interaction regime shows deviations of the Ansatz from the DMC energy curves. Still the agreement remains good for strong and weak interaction strengths, and the qualitative behaviour throughout the crossover from non-interacting to fermionization is captured from the Ansatz. 

\begin{figure*}
\includegraphics[width=12.0 cm,height=8.0cm]{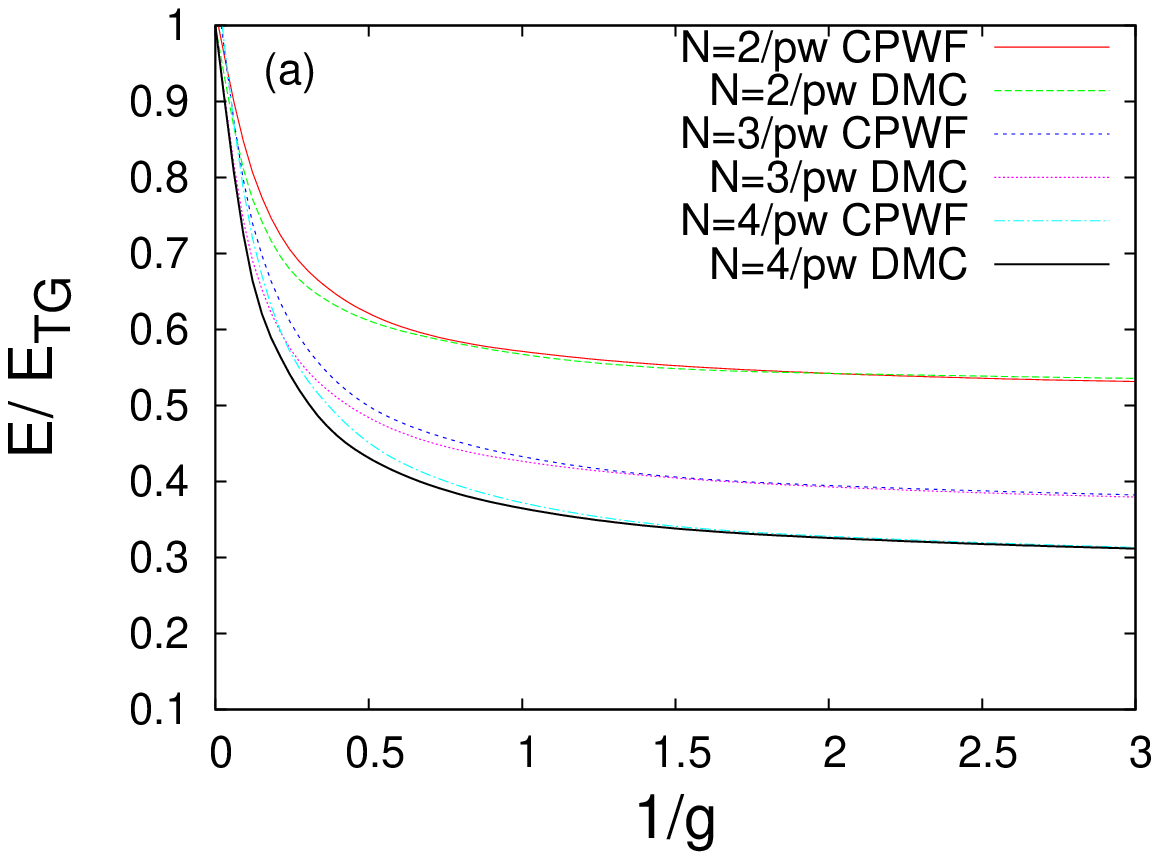}
\includegraphics[width=12.0 cm,height=8.0cm]{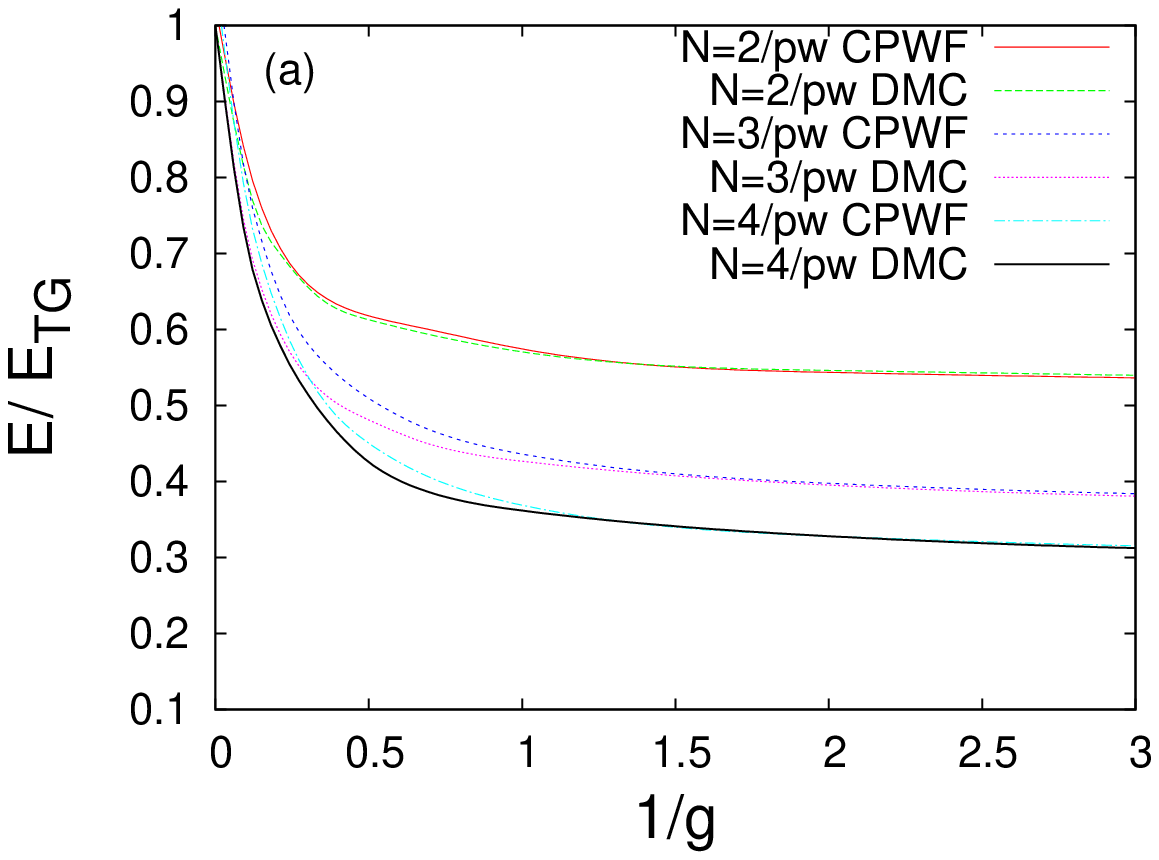}
\caption{  Ground state energies ($E$) divided by the corresponding value of the Tonks-Girardeau limit ($E_{TG}$) for (a) a double and (b) a triple well as a function of the inverse interaction strength ($g=g_{1D}/E_R \alpha$ where $E_R$ is the recoil energy and $\alpha$ the lattice constant). Several cases with respect to the number of atoms per well (pw) are shown, comparing results obtained by DMC and numerical integration of the analytical formula CPWF.}
\label{mw}
\end{figure*}

\begin{figure*}
\includegraphics[width=8.0 cm,height=8.0cm]{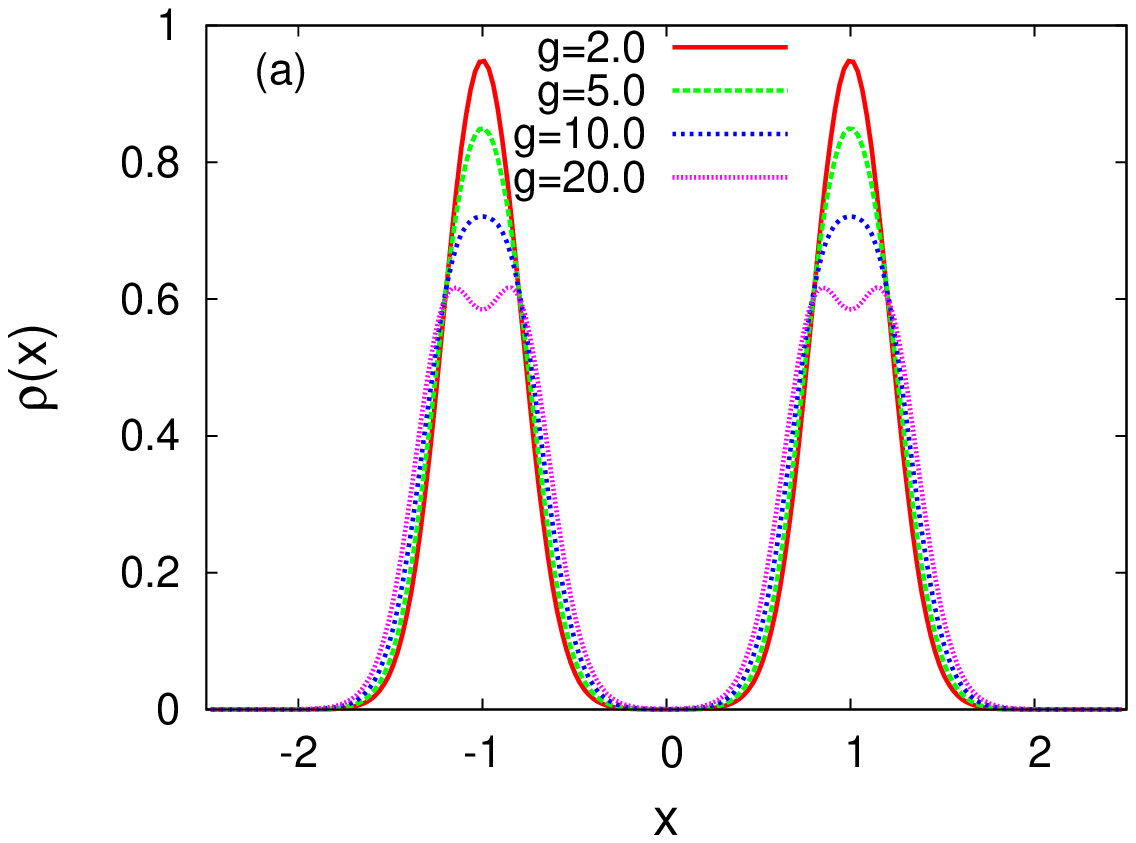} 
\includegraphics[width=8.0 cm,height=8.0cm]{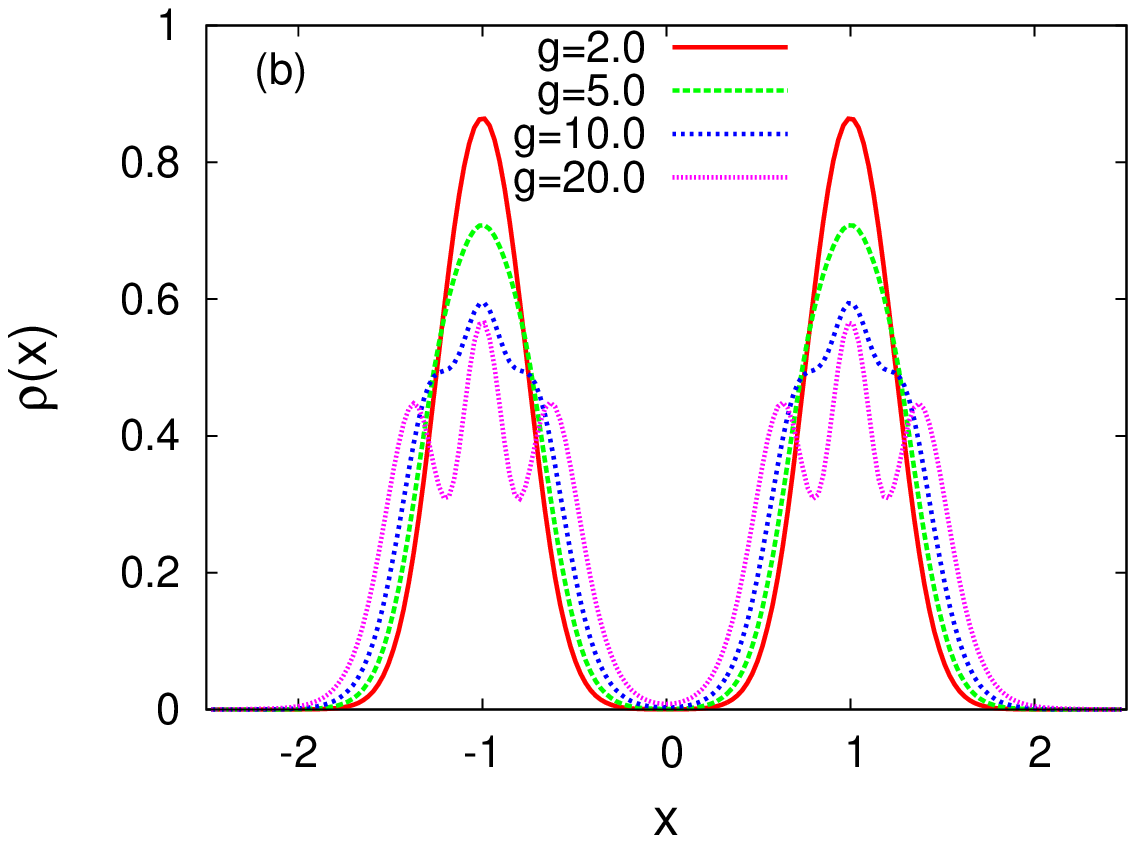} 
\caption{One-body  density functions $\rho(x)$ with the lattice constant $\alpha$ as length scale  for (a) four and (b) six bosons in the double well for several interaction strengths covering the crossover from weak to strong couplings.}
\label{1bd}
\end{figure*}

\begin{figure*}
\includegraphics[width=8.0 cm,height=8.0cm]{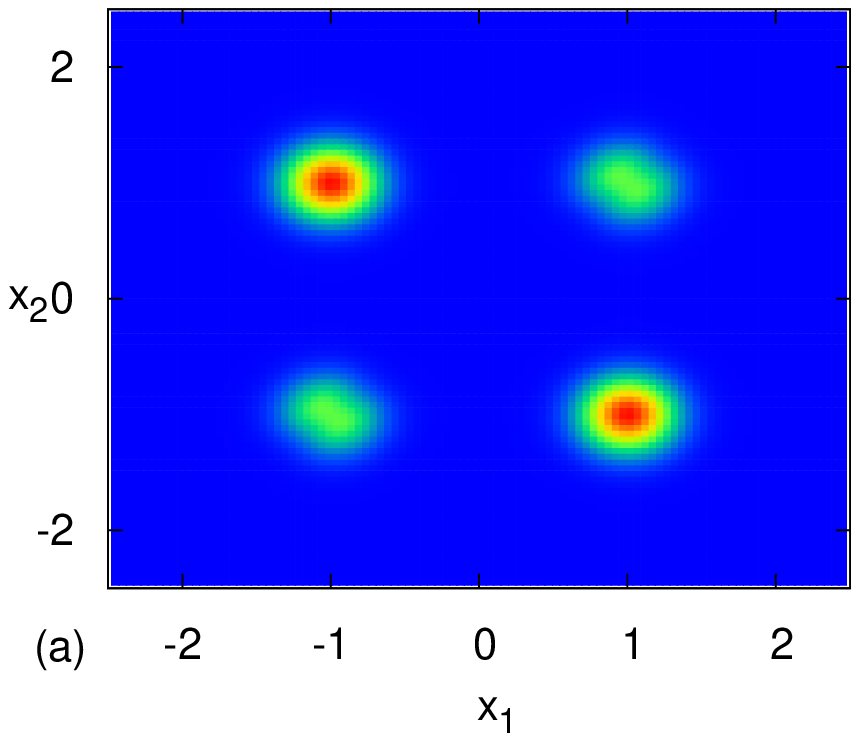} 
\includegraphics[width=8.0 cm,height=8.0cm]{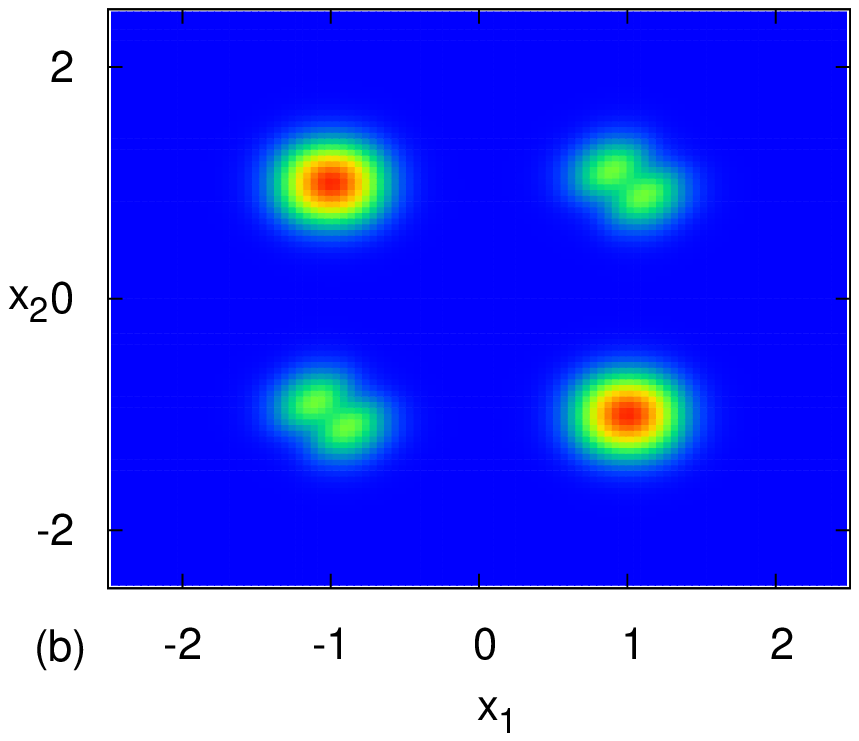} \\
\includegraphics[width=8.0 cm,height=8.0cm]{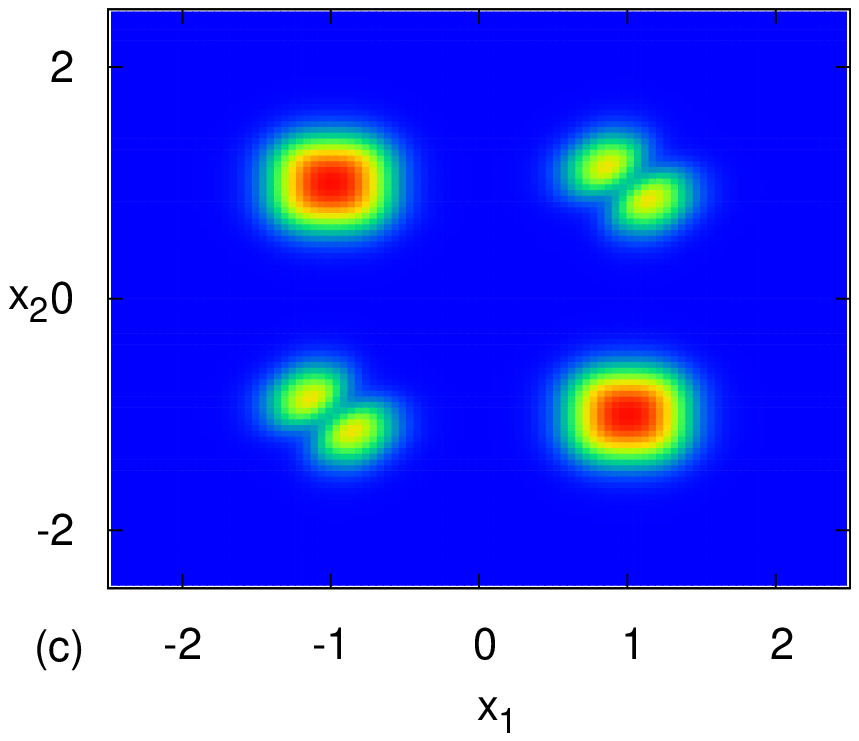} 
\includegraphics[width=8.0 cm,height=8.0cm]{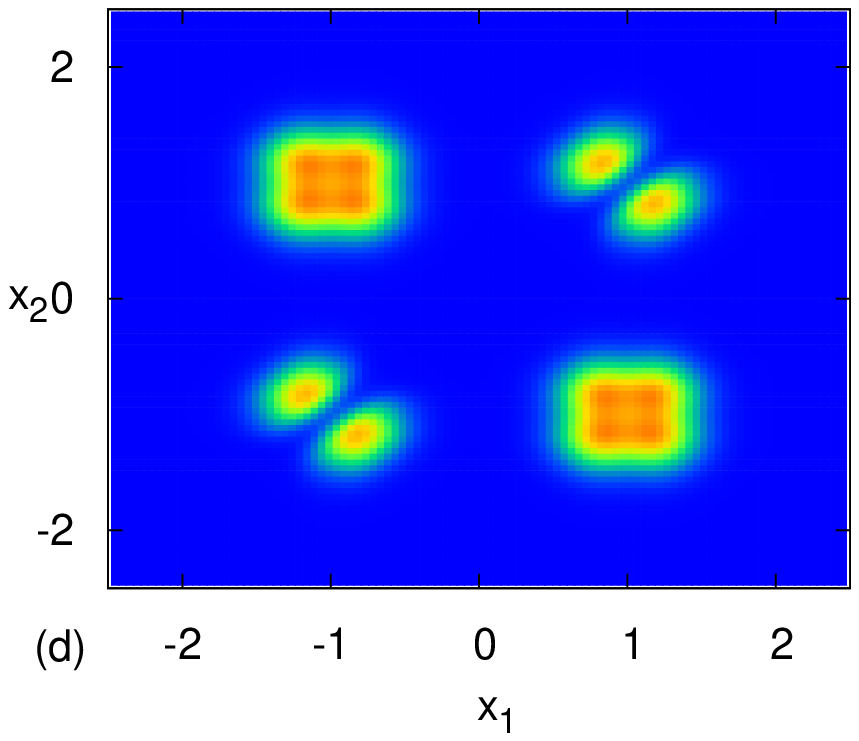} \\
\caption{Two-body density function with  the lattice constant $\alpha$ as length scale  for 4 atoms in the double well for (a) g=2.0 (b) g=5.0 (c) g=10.0 (d) g=20.0}
\label{2bd4}
\end{figure*}

\subsection{Double and triple well}

The generalization of the CPWF to multi-well systems using the corresponding single-particle part (Eq. \ref{spp_lat}), is also compared in the following with DMC results. We consider here cases that are also relevant for cold-atom experiments, i.e.,  two to four particles per well. 

In Fig.~\ref{mw} the curves obtained from the Ansatz and the DMC are presented with respect to the same variables as in Fig.~\ref{ht} for the double [Fig.~\ref{mw} (a)] and the triple [Fig.~\ref{mw} (b)] well potentials. The agreement between energies obtained by numericallly intergrating the CPWF and DMC  calculations is very good for the case of two atoms per site both for the double and the triple well potentials. This is to be expected since the correlated-pair wave function is actually the exact one in the case of a harmonic trap for two atoms \cite{busch,brouzos}. However for three and four atoms per well the deviations start to be substantial at intermediate interaction strength ($g \approx 1$) and become even stronger as we approach the resonance (see Fig.~\ref{mw}). The reason for these deviations lies in the nature of multi-well potentials. As long as the atoms stay in low energy states, which is the case for weak interactions the harmonic approach of each well remains a good approximation.  However, as the interaction strength increases and the atoms come energetically closer to the threshold of the barriers of the potential, each well becomes more and more anharmonic, as well as the tunneling coupling to the neighbouring wells increases. This breakes the validity of a harmonic approach  and produces deviations of the Ansatz which still though works qualitatively and as we will see next captures important properties of the on-site behaviour beyond the weak coupling and the Bose-Hubbard regime. Another important difference compared to the case of the harmonic trap is that here the fermionization limit ($g \to \infty$) is not reproduced exactly by the Ansatz, due to the anharmonicity explained above. Therefore one of the major advantages that our Ansatz proposed in \cite{brouzos} possesses in the harmonic trap, namely reproducing the exact behaviour in the two limiting cases (non- and infinitely interacting ensemble), is missing here. Yet the predictions of the Ansatz are slightly better if one treats the parameter $\beta$ variationally (VMC calculations show a correction up to $1-2\%$ in the error), but certainly there is space for improvement combining probably Bose-Hubbard and exact solutions for the particular profile of the multi-well potential in the analytical description.

\subsection{On-site effects}

In general as shown in several publications \cite{sascha,lattice}, in 1D the density distribution of atoms according to the Bose-Hubbard model, does not capture the plethora of phenomena occuring beyond standard Mott-insulator. Especially for strong interactions interesting on-site effects appear in cases where more than one atom are localized in the same well. The density there may broaden or even acquire two or more peaks due to the repulsion of the atoms on-site (see refs. \cite{lattice}). Our Ansatz here captures these effects due to the two-body part with the correlated-pair wave function \cite{brouzos}. 

 The one-body density function $\rho(x)= \int_{-\infty}^{\infty} ... \int_{-\infty}^{\infty} |\Psi(x,x_2,...,x_N)|^2 dx_2...dx_N$ (with $\Psi(x_1,x_2,...,x_N)$ normalized so that also $\int \rho(x) dx=1$) for the cases (a) four and (b) six atoms in the double well is shown in Fig.~\ref{1bd}. The density funtion at each well becomes broader (see $g=2.0,5.0$) acquires a small plateau ($g=10.0$) and finally develops a number of maxima corresponding to the number of atoms localized per site. Similar behaviour has been observed e.g. in \cite{lattice,sascha} being qualitatively in agreement with our Ansatz. In Fig.~\ref{2bd4} the two-body density  function $\rho_2(x_1,x_2) \equiv \int_{-\infty}^{\infty} ... \int_{-\infty}^{\infty} |\Psi|^2  dx_3...dx_N $ or pair-correlation function  of 4 atoms in the double well is plotted. For weak interactions [$g=2.0$ Fig.~\ref{2bd4} (a)] we have only 4 distinct peaks due to the double well potential. But as the interaction strength increases [$g=5.0$ Fig.~\ref{2bd4} (b)] the diagonal $x_1=x_2$ tends to deplete (the atoms avoid being at the collision points). For stronger interactions  [$g=10.0$ Fig.~\ref{2bd4} (c)] even the off-diagonal peaks broaden  while close to fermionization [$g=20.0$ Fig.~\ref{2bd4} (d)] there are additional peaks appearing at the off-diagonal spots. Similar effects where shown in \cite{sascha, lattice} and are in qualitative agreement with  those obtained by our analytical function.

\section{Conclusions and outlook}

We have studied the crossover from weak to strong interactions of bosonic systems in a harmonic trap and in finite multi-well potentials comparing two approaches: the numerically exact Quantum Monte Carlo method and the correlated-pair wave function approach. Our Ansatz for the harmonic trap introduced and studied in \cite{brouzos} for few-body ensembles, is here tested for much larger systems by comparing the predictions for the energy with Diffusion Monte Carlo calculations. A good agreement all over the interaction regime  but particularly close to the Tonks-Girardeau gas and for very weak interactions is demonstrated. For larger ensembles there exist deviations for the intermediate interaction regime. For finite multi-well systems, namely double and triple wells, we have introduced here a modification of the single-particle part of the Ansatz  in terms of localized functions. We have shown for the energy (different loading of the potential from two to four atoms per well) that there is qualitative agreement but also strong quantitative deviations as we approach the resonance ($g \to \infty$) since the harmonic-approach for each well becomes invalid for energetically excited atoms. Still important effects for strong interactions on the one- and two-body density functions are captured especially concerning on-site features like broadening and appearance of additional peaks. The extension of this idea to fermions or mixtures \cite{brouzos_in_prep} and the improvement of it to describe more accurately the full crossover may represent a valuable input to relevant experimental studies.

\paragraph{Acknowledgements}
The authors are thankful to  Michael Haas and R\"udiger Schmitz for their help on the implementation of the Quantum Monte Carlo code.


\begin{thebibliography}{10}
\bibitem{note}
According to \cite{rey} $f_j$ are the Fourier coefficients of the Mathieu function $ce \left( x, -q \right)$: $f_j= \frac{1}{\pi} \int ce \left( x, -q \right) \cos \left( 2 j x \right) dx$. The parameter $q=\frac{8J}{\omega^2}$  where:
\begin{equation}
\nonumber
J= - \int  w (x_j) \left( -\frac{1}{2} \frac{d^2}{dx^2} + V_0 \sin^2 \left( \pi x /2 \right) \right) w( x_{j-1})  
\end{equation}
characterizes the system, where $w(x)$ are the localized Wannier functions for each well . If Gaussian profile functions are used for w(x) like in Eq. \ref{spp_lat} one obtains 
\begin{equation}
\nonumber
q(V_0)= -\sqrt{8\pi}V_0^{1/4} e^{-2\sqrt{V_0}} \left[ \left( 3 \sqrt{V_0} -1 \right) +e^{-\frac{\pi^2}{8\sqrt{V_0}}}\sqrt{V_0}\right].
\end{equation}
\end{thebibliography}
\end{document}